\documentclass[12pt,preprint]{aastex}
\usepackage{lscape}

\shorttitle{Modeling the Infrared Spectrum of an Unresolved Earth-Moon System}
\shortauthors{Robinson}

\begin{document}

\title{Modeling the Infrared Spectrum of the Earth-Moon System: Implications 
for the Detection and Characterization of Earthlike Extrasolar Planets and 
their Moonlike Companions}

\author{Tyler D. Robinson\altaffilmark{1},\altaffilmark{2}}
\affil{Astronomy Department, University of Washington, Seattle, WA 98195}
\email{robinson@astro.washington.edu}

\altaffiltext{1}{NASA Astrobiology Institute}
\altaffiltext{2}{University of Washington Astrobiology Program}

\begin{abstract}
{\small
Large surface temperatures on the illuminated hemisphere of the Moon 
can lead it to contribute a significant amount of flux to spatially 
unresolved infrared (IR) observations of the Earth-Moon system, 
especially at wavelengths where Earth's atmosphere is absorbing.  We 
have paired the NASA Astrobiology Institute's Virtual Planetary Laboratory 
three-dimensional spectral Earth model with a model of the phase 
dependent IR spectrum of a Moonlike satellite to investigate the 
effects of an unresolved companion on IR observations of Earthlike 
extrasolar planets.  For an extrasolar twin Earth-Moon system observed 
at full phase at IR wavelengths, the Moon consistently comprises about 
20\% of the total signal, approaches 30\% of the signal in the 
9.6~$\mu$m ozone band and the 15~$\mu$m carbon dioxide band, makes 
up as much as 80\% of the total signal in the 6.3~$\mu$m water 
band, and more than 90\% of the signal in the 4.3~$\mu$m carbon 
dioxide band.  These excesses translate to inferred brightness 
temperatures for Earth that are too large by about 20-40 K, and 
demonstrate that the presence of an undetected satellite can have a 
significant impact on the spectroscopic characterization of terrestrial 
exoplanets.  The thermal flux contribution from an airless companion 
depends strongly on the star-planet-observer angle (\emph{i.e.}, 
the phase angle), allowing moons to mimic or mask seasonal variations 
in the host planet's IR spectrum, and implying that observations of 
exoplanets should be taken when the phase angle is as small as 
feasibly possible if contributions from airless companions are to be 
minimized.  We show that, by differencing IR observations of an Earth 
twin with a companion taken at both gibbous phase and at crescent 
phase, Moonlike satellites may be detectable by future exoplanet 
characterization missions for a wide range of system inclinations.
} 
\end{abstract}


\section{Introduction}

The Moon has played a crucial role in maintaining the long term stability 
of Earth's obliquity and, thus, climate \citep{laskaretal93}, although the 
presence of a large satellite does not always guarantee such stability 
\citep{wardetal02}.  Furthermore, simulations indicate that the 
Moon forming impact \citep{hartmannetal86} could have driven away a 
significant mass of volatiles, such as water, from the proto-Earth 
\citep{genda&abe05}.  Thus, the presence of a large moon has important 
consequences for our characterization and understanding of terrestrial 
extrasolar planets.  

Planet formation simulations show that giant impacts like the Moon forming 
impact may be common \citep{idaetal97,canup04,elseretal11}. Consequently, 
moons are likely to contribute to observations of exoplanets, and these 
satellites are likely to be unresolved from their host.  For example, the 
angular separation of the Earth and Moon at a distance of five parsecs is 
smaller than 0.5 mas, which is below the angular resolution of future 
exoplanet characterization missions 
\citep{beichmanetal99,cash06,beichmanetal06,traubetal06}. Recent near 
infrared (IR) observations of the Earth-Moon system from NASA's EPOXI 
mission \citep{livengoodetal11} demonstrated that the Moon can contribute 
a significant amount of the combined flux at wavelengths where Earth's 
atmosphere is strongly absorbing, an effect mentioned in 
\citet{desmaraisetal02}. Thus, an exomoon can affect our understanding of 
its host, be it through clarification or obfuscation, making it prudent to 
investigate how the presence of an exomoon may be detected or inferred, 
and how the presence of an undetected moon could confound observations of 
terrestrial exoplanets. 

To date, ideas for detecting exomoons have focused on transit 
phenomena or bolometric IR lightcurves.  In general, the 
detection of exomoons may be possible through transit 
timing/duration variations and/or transits of the satellite 
across the host planet 
\citep{sartoretti&schneider99,kipping09,sato&asada10}, but 
both scenarios are highly improbable (for example, the 
transit probability of Earth across the Sun is about 1/200, and  
the transit probability of the Moon across Earth's visible disk 
is about 1/50).  Target star lists for future exoplanet 
characterization missions usually contain about 100 stars 
\citep{beichmanetal99,brown05}, which implies that such missions 
will likely not detect exomoons using transit effects.  
Mutual shadowing of a planet and satellite can reveal the 
presence of a moon and has a high geometric probability of 
occurring \citep{cabrera&schneider07}.  Unfortunately, 
detecting such an event would require a duty cycle of nearly 
100\% of the moon's orbit, which, for the Earth-Moon system, 
would consume nearly a month of continuous monitoring.  It 
may be possible to detect Earth-sized satellites with distinct 
atmospheres from their host planet in reflected light at 
wavelengths where the host is strongly absorbing 
\citep{williams&knacke04}, but more work is required to quantify 
the effect and to identify techniques for discrimination.

\citet{moskovitzetal09} investigated the effects of airless  
satellites on IR lightcurves of planets similar to Earth  
(\emph{i.e.}, Earthlike planets) using an energy balance model 
to simulate climate on the host and bolometric thermal flux 
models to simulate observations \citep{gaidos&williams04}. 
The contribution of the airless companion depends on phase and, 
thus, varies smoothly over an orbital period, which causes the 
satellite's emission to mimic seasonal variations in the planet's 
emitted thermal flux.  These authors found that the detection of a 
satellite around an Earthlike planet is only feasible for Mars-sized 
moons, and that the nondetection of a satellite can lead to the 
mischaracterization of the host planet's obliquity, orbital longitude of 
vernal equinox relative to inferior conjunction, and thermal properties. 
However, direct IR observations of exoplanets will likely be spectrally 
resolved, not bolometric.  Thus, it is important to investigate how the 
addition of spectral resolution will change these conclusions.

In this paper we use spectrally resolved models to determine  
the significance of the Moon's contribution to spatially 
unresolved IR observations of the Earth-Moon system, and the extent to 
which a Moon twin (or exoMoon) may influence spectroscopic characterization 
of an Earth twin (or exoEarth).  We investigate how the thermal flux 
from a world similar to the Moon (\emph{i.e.}, a Moonlike world), and its 
phase dependence, can be used to detect moons orbiting Earthlike exoplanets.  
Finally, we discuss the implications of our findings for future exoplanet 
detection and characterization missions.

\section{Model Description}

Our models of the disk integrated spectra of Earth and a Moonlike world 
compute the integral of the projected area weighted intensity in 
the direction of the observer, which can be written as 
\begin{equation}
F_{\lambda}\left({\bf \hat{o}},{\bf \hat{s}}\right) = \frac{R^{2}}{d^{2}}\int I_{\lambda}\left({\bf \hat{n}},{\bf \hat{o}},{\bf \hat{s}}\right) \left({\bf \hat{n}} \cdot {\bf \hat{o}}\right)d\omega \ ,
\label{eqn:fluxint}
\end{equation}
where $F_{\lambda}$ is the disk integrated specific flux density received 
from a world of radius $R$ at a distance $d$ from the observer,   
$I_{\lambda}\left({\bf \hat{n}},{\bf \hat{o}},{\bf \hat{s}}\right)$ is the 
location dependent specific intensity in the direction of the observer, 
${\bf \hat{n}}$ is a surface normal unit vector, ${\bf \hat{o}}$ 
and ${\bf \hat{s}}$ are unit vectors in the direction of the observer and Sun 
(or host star), respectively, and $d\omega$ is an infinitesimally small unit 
of solid angle on the globe. The integral in Eq. \ref{eqn:fluxint} is over the 
entire observable hemisphere ($2\pi$ steradians) and the dot product at the end 
of the expression ensures that an element of area $R^{2}d\omega$ near the 
limb is weighted less than an element of equal size near the sub-observer 
point.  Note that, for reflected light, $I_{\lambda}$ will be zero at 
locations on the night side of the world (\emph{i.e.}, where 
${\bf \hat{n}} \cdot {\bf \hat{s}} < 0$), but is non-zero at all locations 
when considering thermal emission.  The following subsections describe our 
techniques for solving Eq.~\ref{eqn:fluxint} for Earth and a Moonlike companion.

\subsection{Earth Model}

To simulate Earth's appearance to a distant observer, we use the NASA Astrobiology 
Institute's Virtual Planetary Laboratory three-dimensional spectral Earth model, 
which generates temporally and spectrally resolved disk integrated synthetic observations   
of Earth. This model has been described and extensively validated both for temporal 
variability and for a variety of phases, at wavelengths from the near-ultraviolet through 
the IR in previous papers \citep{robinsonetal2010,robinsonetal2011}, so only a brief 
description of the model will be presented here.  

In our simulations, we divide Earth into a number of equal area pixels 
according to the HEALPix scheme \citep{gorskietal05}, thus converting the 
integral in Eq.~\ref{eqn:fluxint} to a sum over the observable pixels.  The 
wavelength dependent intensity coming from any given pixel is assembled from 
a lookup table that contains spectra generated over a grid of different 
solar and observer zenith and azimuth angles. Elements within the lookup table, 
which are generated using a one-dimensional, line-by-line radiative transfer 
model \citep{meadows&crisp96}, are computed for a variety of different surface 
and atmospheric conditions, as well as several different cloud coverage scenarios 
(\emph{e.g.}, thick, low cloud or thin, high cloud).

To simulate time dependent changes in Earth's spectrum we use spatially 
resolved, date specific observations of key surface and atmospheric 
properties from Earth observing satellites as input to our Earth model. 
Gas mixing ratio and/or temperature profiles are taken from the Microwave Limb 
Sounder \citep{watersetal06}, the Tropospheric Emission Spectrometer 
\citep{beeretal01}, the Atmospheric Infrared Sounder \citep{aumann03}, and the 
CarbonTracker project \citep{petersetal07}.  Snow cover and sea ice data as 
well as cloud cover and optical thickness data are taken from the Moderate 
Resolution Imaging Spectroradiometer instruments \citep{salomonsonetal89} aboard 
NASA's Terra and Aqua satellites \citep{halletal95,riggsetal95}. Wavelength 
dependent optical properties for liquid water clouds were derived using a Mie 
theory model \citep{crisp97} and were parametrized using geometric optics for 
ice clouds \citep{muinonenetal89}.

\subsection{Moon Model}

Day side temperatures on the Moon are predominantly determined by the 
radiative equilibrium established between absorbed solar radiation and 
emitted thermal radiation \citep{lawsonetal00}.  Thus, the temperature, 
$T$, at any location on the sunlit portion of the Moon is given by
\begin{equation}
T\left({\bf \hat{n}},{\bf \hat{s}}\right) = \left(\left({\bf \hat{n}} \cdot {\bf \hat{s}}\right)\frac{1-A}{\sigma\epsilon}S\right)^{1/4} \ ,
\label{eqn:temp}
\end{equation}
where $A$ is the surface Bond albedo, $\sigma$ is the Stefan-Boltzmann 
constant, $\epsilon$ is the bolometric emissivity of the surface, and 
$S$ is the bolometric solar flux density at the Moon's 
orbital distance from the Sun (\emph{i.e.}, 1~AU).  Since we are focusing on 
spatially unresolved observations in this study, we assume that $A$ 
and $\epsilon$ do not vary with location on the Moon, using standard 
globally averaged values of $0.127$ and $0.95$, respectively 
\citep{racca95}.  

Assuming a spatially non-varying Bond albedo and bolometric 
emissivity allows us to use Eq.~\ref{eqn:fluxint} to write  
the disk integrated thermal flux density from the Moon, 
$F_{\lambda,M}$, as a function of the star-Moon-observer 
angle (\emph{i.e.}, the phase angle), $\alpha$.  Thus, 
\begin{equation}
F_{\lambda,M} = \frac{R_{M}^{2}}{d^{2}} \bigg[ \int_{\alpha-\pi/2}^{\pi/2} \int_{-\pi/2}^{\pi/2} \epsilon_{\lambda}B_{\lambda}\left(\theta,\phi\right) \cos\theta \cos^{2}\phi d\phi d\theta  + \frac{\pi}{2}\epsilon_{\lambda}B_{\lambda}\left(T_{n}\right)\left( 1-\cos\alpha \right) \bigg] \ ,
\label{eqn:moonflux}
\end{equation}
where $R_{M}$ is the radius of the Moon, $\epsilon_{\lambda}$ is 
the wavelength dependent, global average surface emissivity, 
$B_{\lambda}$ is the Planck function, and $T_{n}$ is the lunar 
nightside temperature.  Note that  
$\cos\alpha = {\bf \hat{o}} \cdot {\bf \hat{s}}$ and, following 
\citet{sobolev75}, we can write ${\bf \hat{n}} \cdot {\bf \hat{s}}$ 
as $ \cos\left(\alpha-\theta\right)\cos\phi$.  We take 
$\epsilon_{\lambda}$ to be an admixture of 17\% lunar mare material 
and 83\% lunar highland material, whose emissivity spectra were 
measured from Apollo lunar samples and taken from the ASTER Spectral 
Library (http://speclib.jpl.nasa.gov/).  Note that we can generalize 
Eq.~\ref{eqn:moonflux} to Moonlike companions, which we take to be 
similar to the Moon in all ways except size, by varying the value 
of $R_{M}$.  

The lunar nightside temperature is measured to be roughly $100$~K 
\citep{racca95}, but our model is not sensitive to the specific 
value that we choose for the nightside temperature since the wavelength 
dependent thermal flux coming from such a cold blackbody is more than 100 
times smaller than the thermal flux coming from Earth or the full phase Moon.  
As noted by \citet{moskovitzetal09}, a large day-night temperature contrast 
for a Moonlike body can be maintained as long as its rotational period is 
above a certain threshold.  For the average lunar surface heat capacity 
and temperature, this timescale is about 20~hours. Longer rotational 
periods than this are likely for Moonlike companions to extrasolar 
Earthlike planets as the timescale for synchronous rotation due to tidal 
forces is small when compared to the lifetime of a low mass 
star \citep{gladmanetal96}.

Our model does not include a phase dependent correction to the Moon's 
thermal flux that is sometimes incorporated into parameterized spectral 
models of airless bodies to account for the so-called beaming effect.  The  
effect amounts to corrections at roughly the 10\% level or less 
\citep{morrison73,mendell&lebofsky82,lebofskyetal86,rozitis&green11}, 
which is small enough to be ignored for this study.  Note that 
Eq.~\ref{eqn:moonflux} does not include a reflected solar component. We 
investigated the importance of reflected sunlight in our results by using the 
EPOXI lunar observations \citep{livengoodetal11} and assuming a Lambert phase 
function to extrapolate the observations to different phases, and found no 
significant change in detectability.  Also, note that by integrating 
Eq.~\ref{eqn:moonflux} over wavelength to produce an analytic expression 
for the bolometric thermal flux, we were able to reproduce the bolometric 
IR lightcurves for the Moon from \citet{moskovitzetal09}.

\section{Results}

\begin{figure}[ht]
  \centering
  {
    \includegraphics[width=2.1in,angle=0]{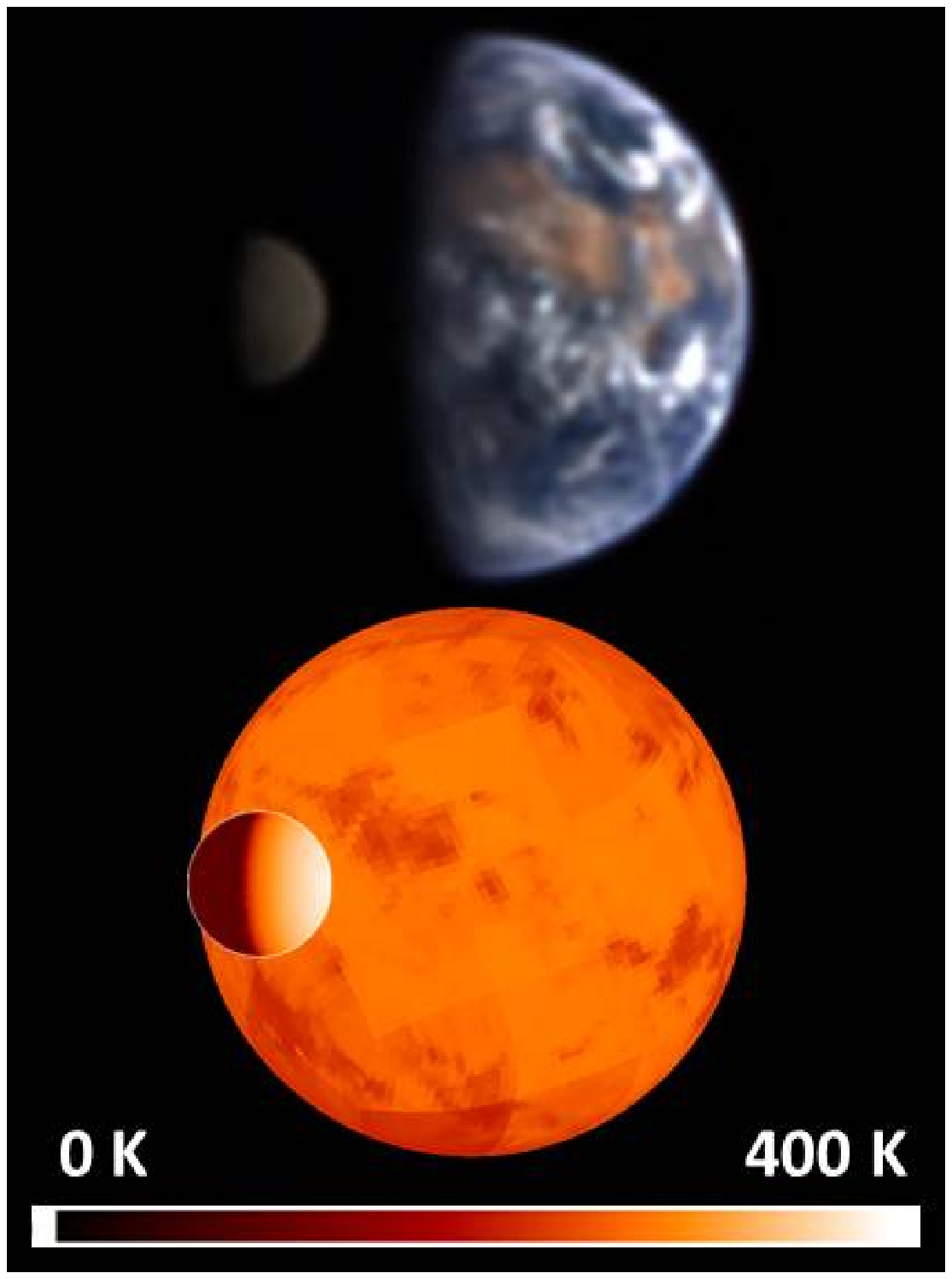}
  }
  {
    \includegraphics[width=4.0in,angle=0]{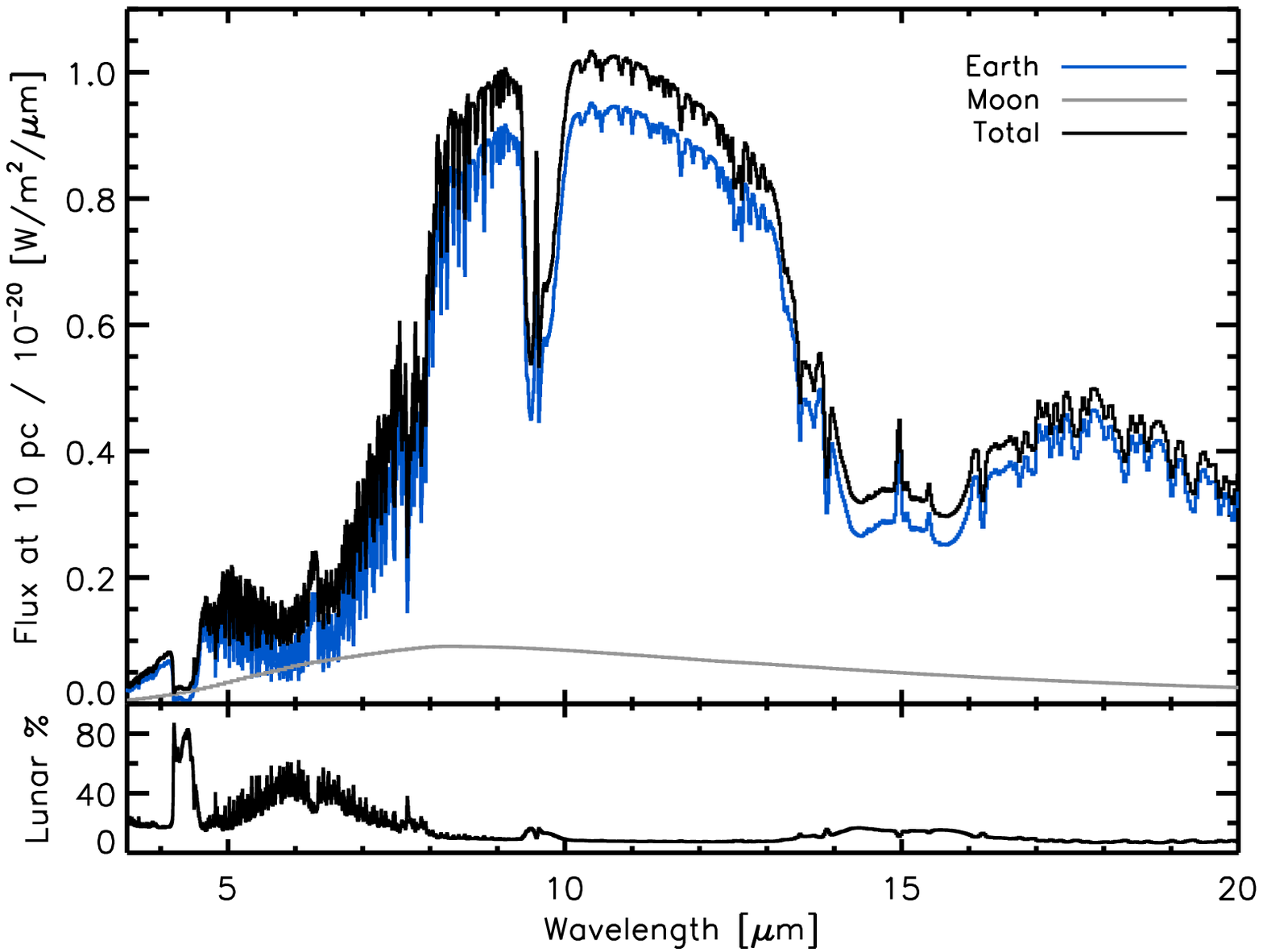}
  }
  \caption{\footnotesize
           True color image of the Earth-Moon system, taken as part of 
           NASA's EPOXI mission compared to a simulated image using 
           10~$\mu$m brightness temperatures from our models. The 
           spectra  on the right shows the corresponding flux at 10~pc 
           from the Moon (grey), Earth (blue), and the combined 
           Earth-Moon flux (black), not including transit effects.  The 
           panel below the spectra shows the wavelength dependent lunar 
           fraction of the total signal.  Images and spectra are for a 
           phase angle of $75.1^{\circ}$}
  \label{fig:epoxi}
\end{figure}

Temperatures near the sub-solar point on the Moon reach nearly 400~K. 
As a result, at thermal wavelengths the brightness of some regions 
on the Moon can be much greater than any region on Earth. In 
Fig.~\ref{fig:epoxi}, we demonstrate this behavior by comparing 
a visible light, true color image from NASA's EPOXI mission, taken 
at a phase angle of $75.1^{\circ}$, and the same image in 
10~$\mu$m brightness temperatures from our Earth and 
Moon models.  Note that intensities from the Moon are quite small in 
the true color visible image, which is due to the relatively low average 
visible albedo of the Moon (about 7\%, compare to about 30\% for Earth).
In the thermal image, though, regions near the sub-solar point on the Moon 
appear brighter than any regions on Earth's disk.  Also shown in 
Fig.~\ref{fig:epoxi} is the corresponding disk integrated flux 
received at 10~pc for the Moon, Earth, and the combined system.  As might be 
expected, the disk integrated Earth significantly outshines the disk integrated 
Moon, with the Moon typically accounting for less than 10\% of the combined 
flux at most IR wavelengths.  However, the Moon contributes as much as 50\% 
of the flux at wavelengths near the 6.3~$\mu$m water band.  The following 
subsections explore the lunar contribution to IR observations of the 
Earth-Moon system and, furthermore, how the wavelength and phase dependent 
nature of this contribution can be used to detect Moonlike satellites around 
terrestrial exoplanets.

\subsection{Lunar Contribution to Combined Flux}

To investigate the extent to which an exoMoon could influence measurements 
of the disk integrated spectrum of an exoEarth, we ran our Earth model for 
a variety of different dates (vernal equinox, as well as mid-northern 
summer and winter) in 2008 (the most recent year for which CarbonTracker 
data were available).  Seasonal variability in disk integrated fluxes from 
Earth were roughly 10-15\% in the 10-12~$\mu$m window region and were 
generally much smaller at other IR wavelengths, which agrees with the 
observations published by \citet{heartyetal09}.  

\begin{figure}[ht]
  \centering
   {
     \includegraphics[width=5in,angle=0]{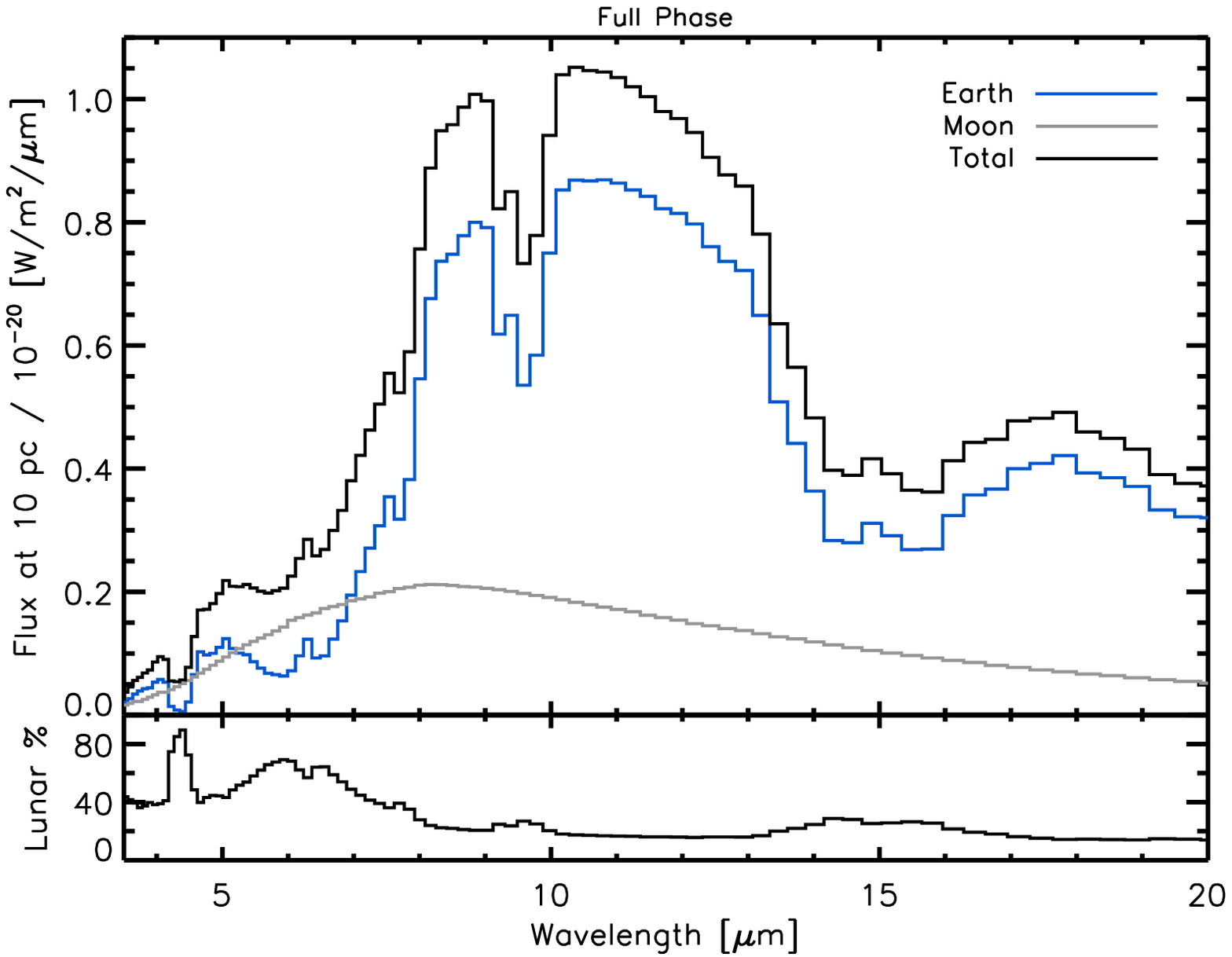}
   }
   { 
     \includegraphics[width=5in,angle=0]{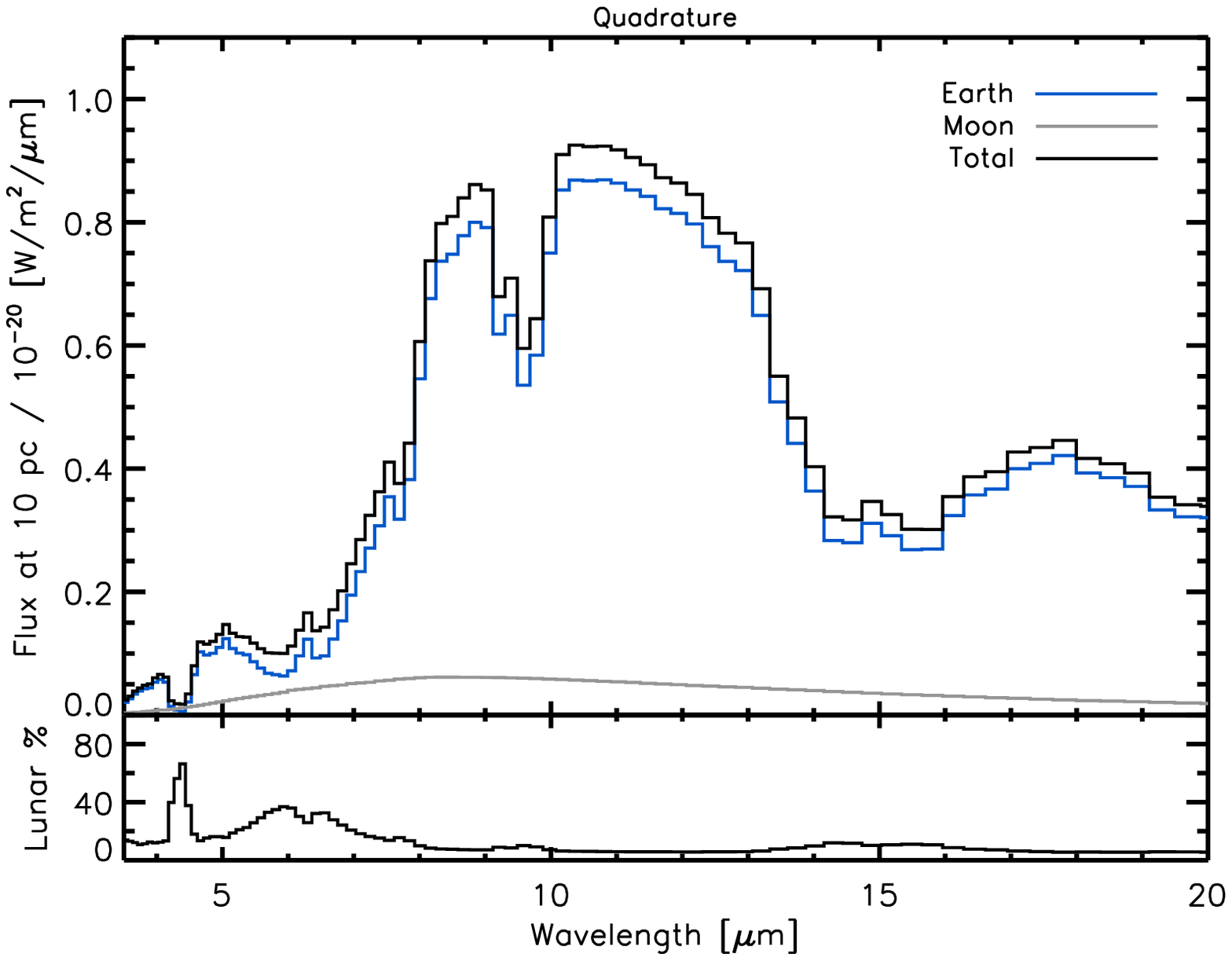}
   }
  \caption{\footnotesize
           Infrared spectra of the Moon (grey), Earth (blue), and the 
           combined Earth-Moon system (black) at full phase (top) and 
           quadrature (bottom).  Spectra are averaged over 24~hours at 
           Earth's vernal equinox, and the spectral resolution is 50.  
           The panels below the spectra show the wavelength dependent 
           lunar fraction of the total signal.}
  \label{fig:examples}
\end{figure}

Figure~\ref{fig:examples} shows the fluxes received from 
the Moon, Earth, and the combined Earth-Moon system at two 
different viewing geometries; full phase and quadrature 
(50\% illumination, phase angle of $90^{\circ}$). In both 
cases, the observations are averaged over 24 hours at 
Earth's vernal equinox, and the observer is viewing Earth's 
orbit edge on and is located over the Equator.  The spectral 
resolution ($\lambda/\Delta\lambda$) in this figure, and all 
subsequent figures, is taken to be 50, which is consistent 
with the resolution for an IR exoplanet characterization 
mission \citep{beichmanetal06}. In the quadrature case, the 
Moon contributes less than 10\% of the net flux from the 
combined Earth-Moon system at most wavelengths, but contributes 
nearly 40\% of the flux within the 6.3~$\mu$m water band and as 
much as 60\% of the flux in the 4.3~$\mu$m carbon dioxide band.

\begin{figure}[ht]
  \centering
  \includegraphics[width=5in,angle=0]{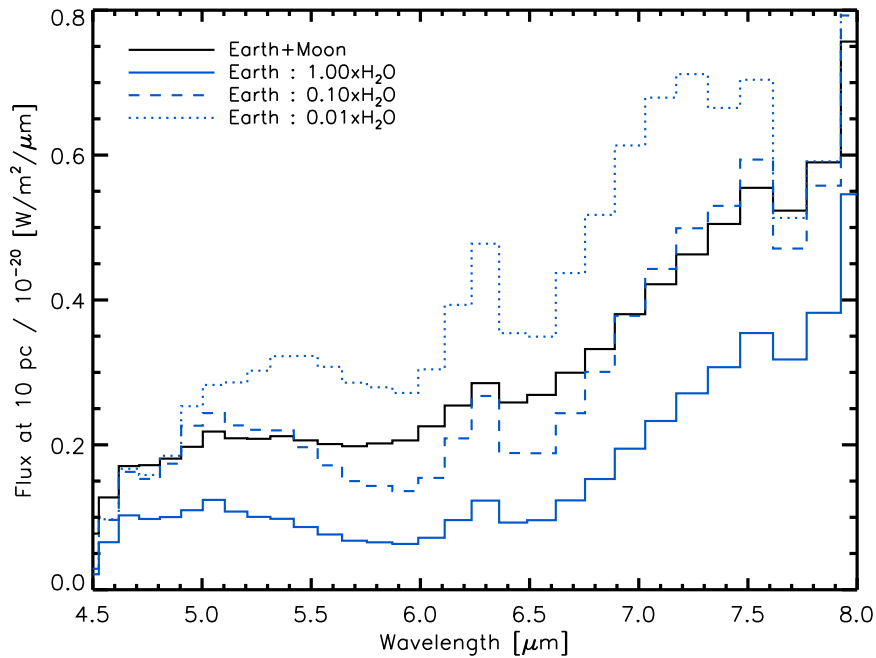}
  \caption{\footnotesize
           Earth's 6.3~$\mu$m water band with and without the full phase flux from the 
           Moon (black and solid blue, respectively, from Fig.~\ref{fig:examples}).  Also 
           shown are IR spectra of Earth with artificially lowered amounts of water vapor 
           in the atmosphere generated using a one-dimensional, line-by-line radiative transfer 
           model \citep{meadows&crisp96}.  The dashed blue line shows the case where water vapor 
           mixing ratios are at 10\% their present day levels and the dotted blue line is 
           for 1\% present day levels.  The addition of the Moon's flux fills in the water 
           absorption feature, causing the feature to more closely mimic an Earth with roughly 
           90\% less water vapor in the atmosphere.}
  \label{fig:watervarn}
\end{figure}

The full phase case presented in Fig.~\ref{fig:examples} 
shows that it is possible for the Moon to contribute a 
significant amount of flux to combined Earth-Moon 
observations.  In this scenario, the lunar thermal radiation 
consistently comprises about 20\% of the total signal, 
approaches 30\% of the signal in the 9.6~$\mu$m ozone 
band and the 15~$\mu$m carbon dioxide band, makes up 
as much as 80\% of the total signal in the 6.3~$\mu$m 
water band, and exceeds 90\% of the signal in the 4.3~$\mu$m 
carbon dioxide band.  The added flux within the water band 
causes the feature to more closely resemble a spectrum of 
Earth with water vapor mixing ratios artificially lowered 
to 10\% of their present level, creating the appearance of 
a much drier planet.  This effect is demonstrated in 
Fig.~\ref{fig:watervarn}, where we show the 6.3~$\mu$m water 
band from the full phase case in Fig.~\ref{fig:examples} 
along with spectra in which Earth's water vapor mixing 
ratios have been artificially scaled to 10\% and 1\% of 
their present day levels.

\begin{figure}[ht]
  \centering
  \includegraphics[width=5in,angle=0]{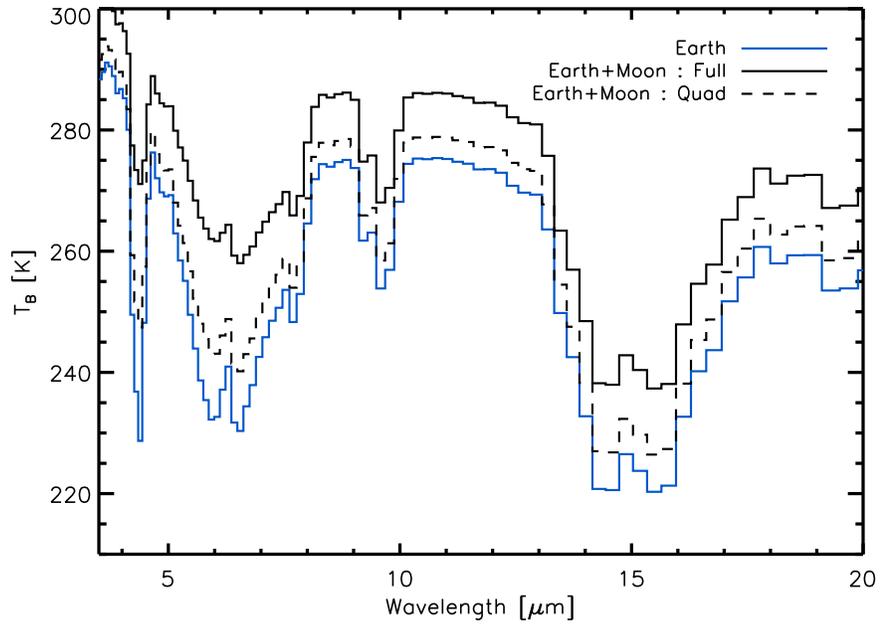}
  \caption{\footnotesize
           Brightness temperature spectra for Earth (blue) as well as 
           for the combined Earth-Moon system at full phase (black, solid) 
           and at quadrature (black, dashed).  Brightness temperatures 
           were calculated using the corresponding fluxes from  
           Fig.~\ref{fig:examples}, one Earth radius was used in the 
           conversion from flux to intensity, and the spectral resolution 
           is 50.}
  \label{fig:brightnesstemp}
\end{figure}

Figure \ref{fig:brightnesstemp} shows how the additional 
contribution from the Moon in the full phase and quadrature 
cases could confuse brightness temperature measurements and, 
thus, characterization attempts.  At quadrature, brightness 
temperatures are increased by 5-10~K in the 6.3~$\mu$m water 
band and the 15~$\mu$m carbon dioxide band, and by about 15~K 
in the 4.3~$\mu$m carbon dioxide band, as compared to those 
expected from Earth alone. At full phase, temperatures 
measured in the 15~$\mu$m carbon dioxide band are about 20~K 
above those expected for Earth alone and, strikingly, temperatures 
measured in the 4.3~$\mu$m carbon dioxide band and the 
6.3~$\mu$m water band are as much as 30-40~K larger. In the 
window region, located between the 9.6~$\mu$m ozone band and the 
15~$\mu$m carbon dioxide band, where Earth's atmosphere is 
relatively free of gaseous absorption and, thus, brightness 
temperatures measurements are more ideal for surface temperature 
retrievals, the Moon increases temperature measurements 
by about 5~K in the quadrature case and more than 10~K 
in the full phase case.

Clear differences can be seen in Fig.~\ref{fig:brightnesstemp} 
between the Earth-only spectrum as compared to Earth-Moon system 
spectra in both the shapes and depths of absorption features. 
The first indications that an undetected exomoon may be 
orbiting a directly imaged exoplanet may come from such 
discrepancies.  For example, the 6.3~$\mu$m water band is 
quite symmetric about its center in the Earth-only spectrum 
(at wavelengths shortward of the 7.7~$\mu$m methane band), but 
the feature appears strongly asymmetric when the flux from the 
full phase Moon is added.  Furthermore, the bases of the 
4.3~$\mu$m and 15~$\mu$m carbon dioxide features are sensitive 
to similar pressure levels in Earth's atmosphere and, thus, return 
similar brightness temperatures in the Earth-only spectrum.  
When the thermal flux from the full phase or quadrature Moon 
is added, though, the temperatures recorded in the 4.3~$\mu$m 
band are greatly increased, leading to a discrepant appearance 
between the two carbon dioxide bands.

\subsection{Detecting Exomoons via Phase Differencing}

The thermal flux from a slowly rotating, airless companion depends 
strongly on phase angle (Eq.~\ref{eqn:moonflux}). As a result, an exomoon 
can present a time varying signature with a period equal to the host's 
orbital period which can be masked by (or mimic) any seasonally 
dependent thermal flux variations from the host planet.  However, 
the phase dependent contribution from an exomoon may be detectable by 
differencing IR observations taken at two different phase angles at 
wavelengths where the moon is relatively bright and the host planet's 
spectrum exhibits only small seasonal variations.

\begin{figure}[ht]
  \centering
  {
    \includegraphics[width=3.125in,angle=0]{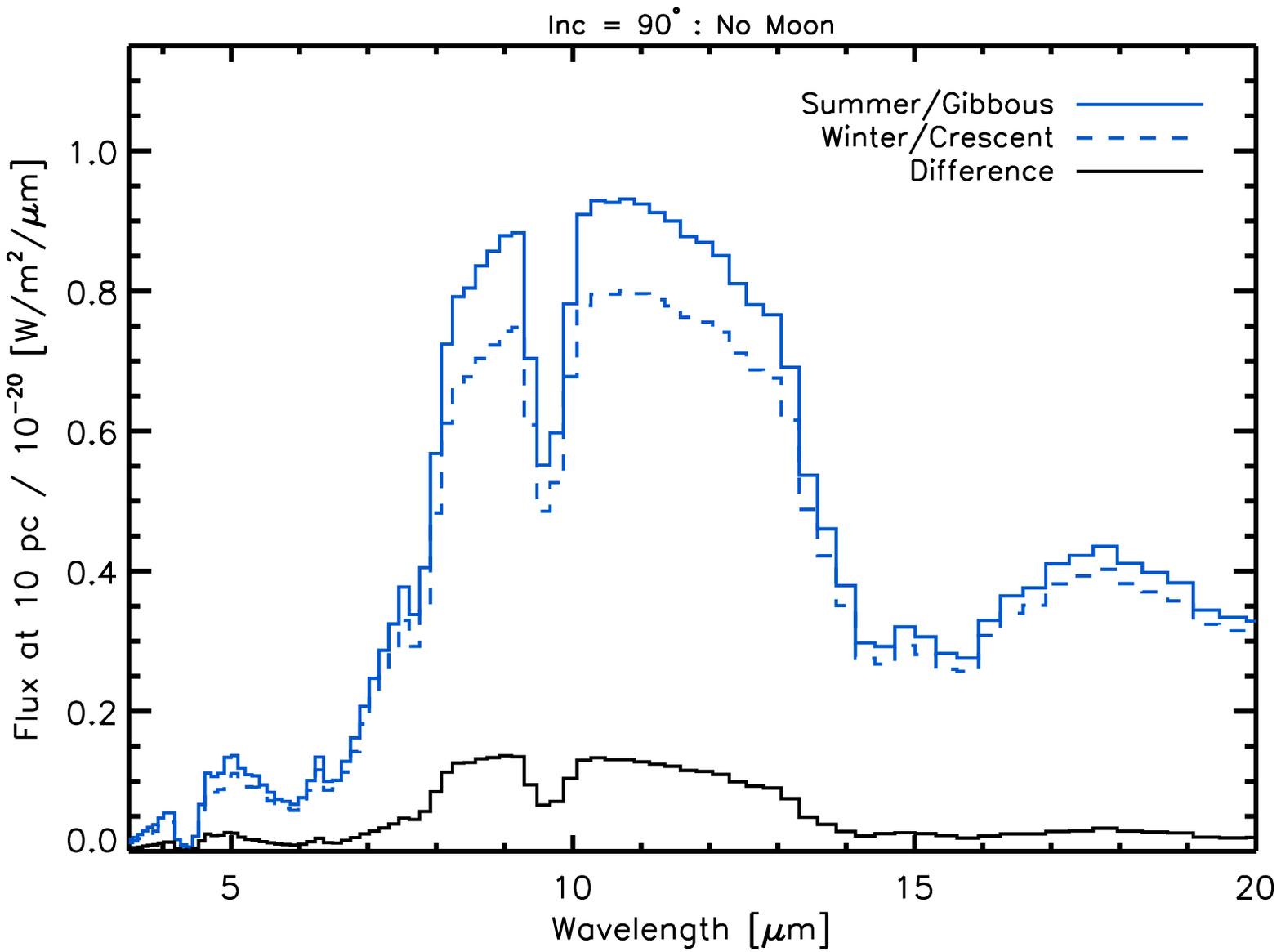}
  }
  { 
    \includegraphics[width=3.125in,angle=0]{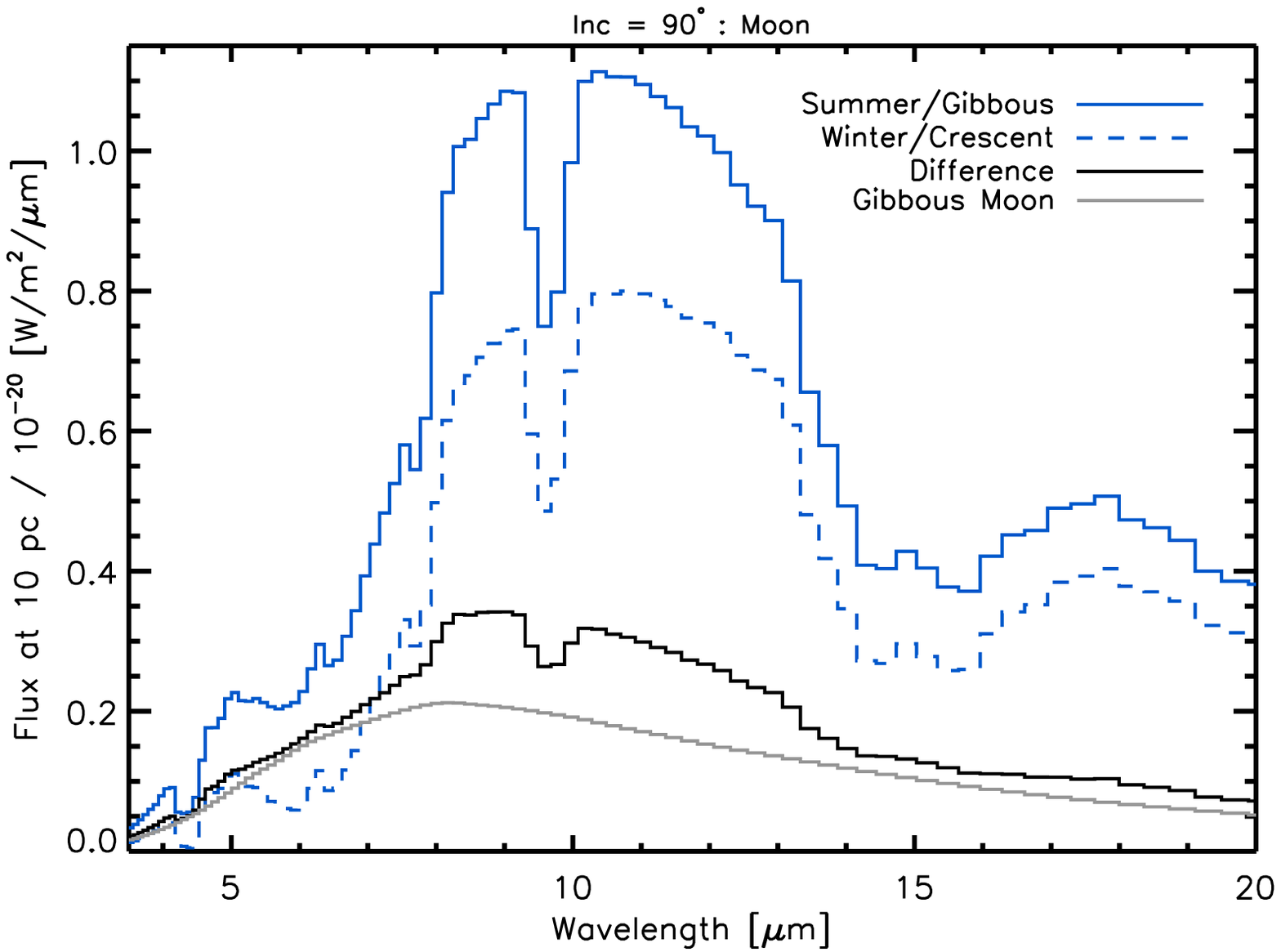}
  }
  {
    \includegraphics[width=3.125in,angle=0]{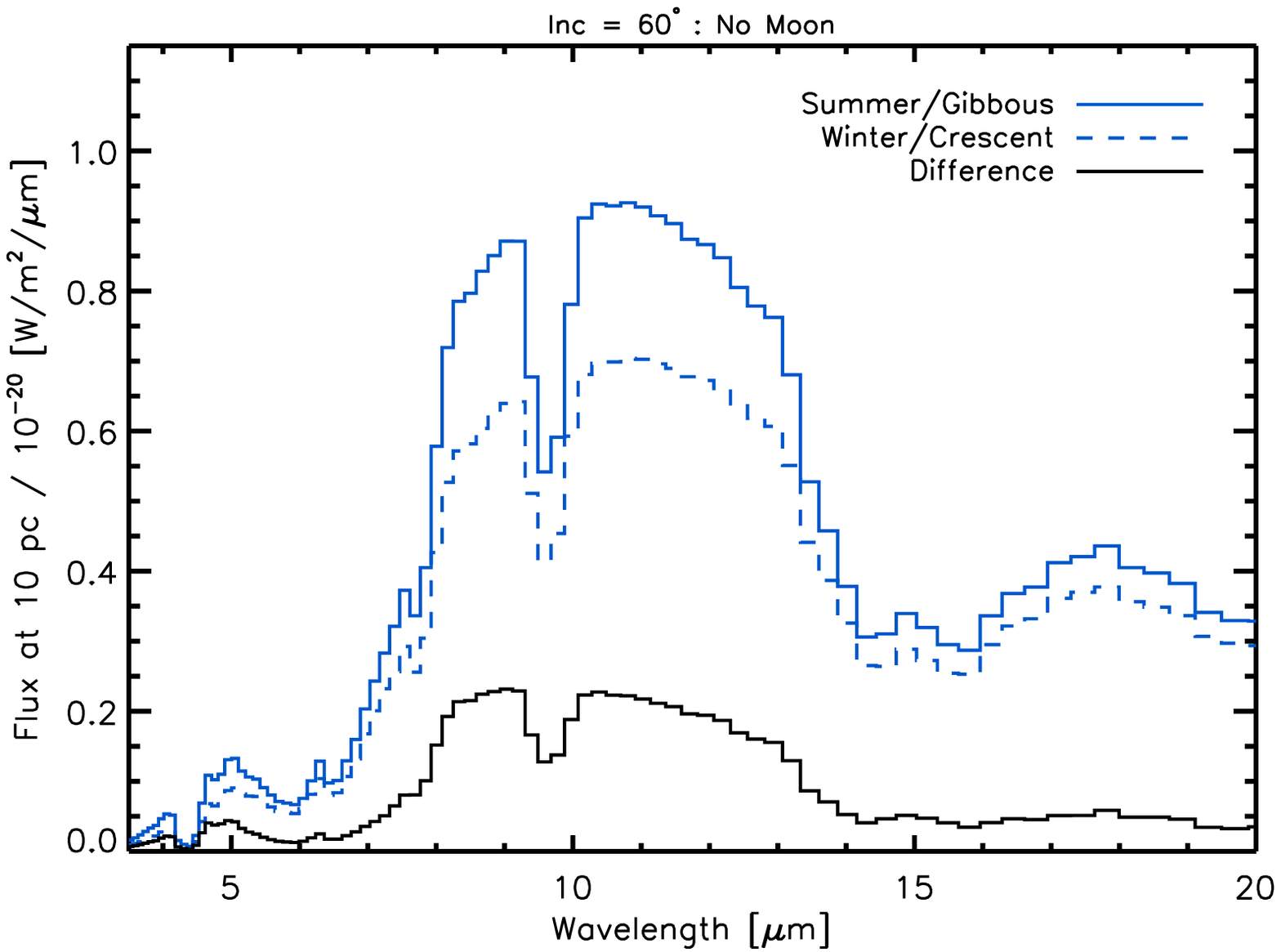}
  }
  { 
    \includegraphics[width=3.125in,angle=0]{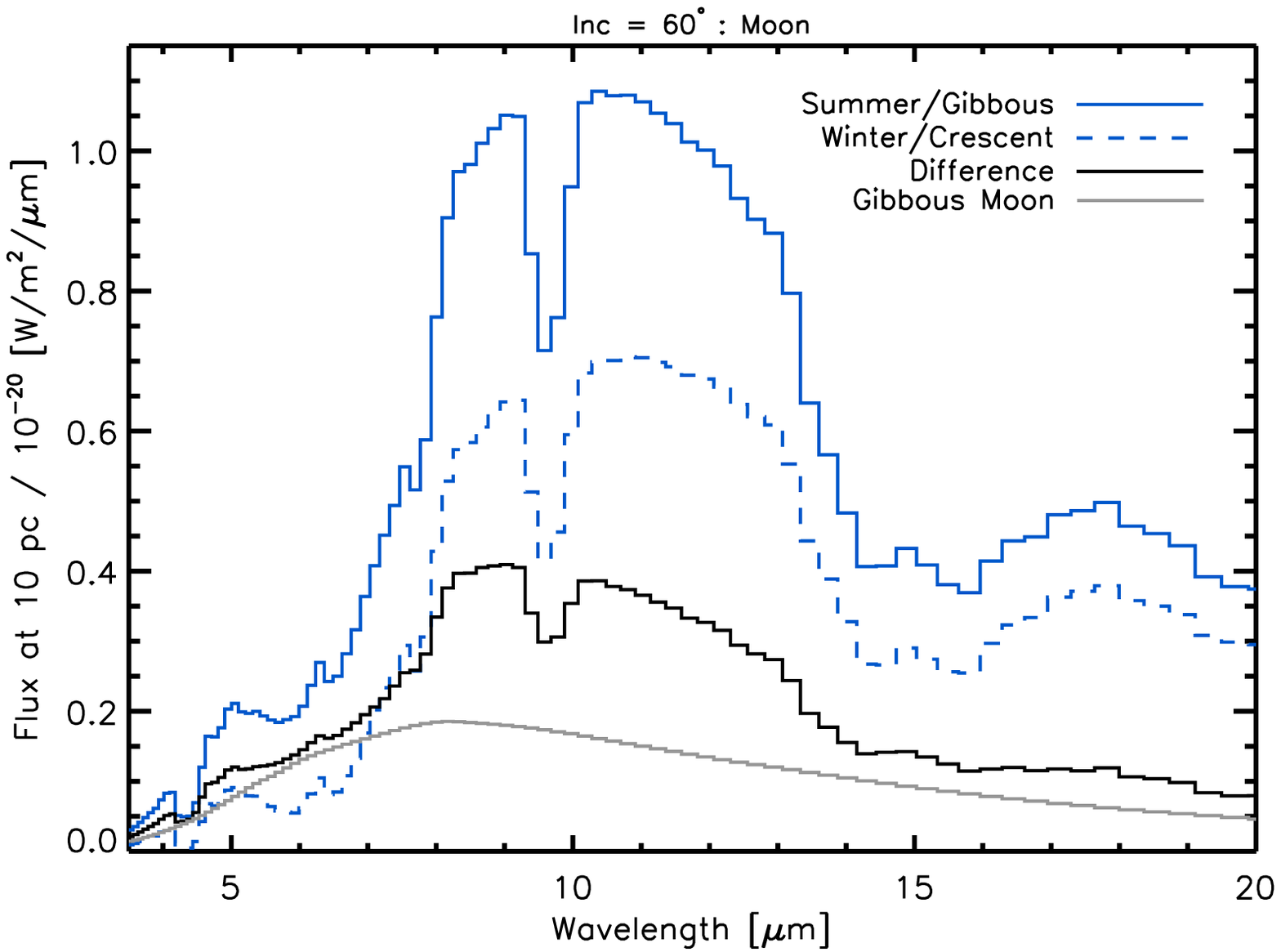}
  }
  \caption{\footnotesize
           Simulated observations of both an exoEarth (left column) as well 
           as an exoEarth-Moon system (right column), both at a distance of 
           10~pc, demonstrating the phase differencing technique which could 
           be used to detect exomoons. Observations are averaged over 24~hours 
           and the spectral resolution is 50. One observation is taken at a small 
           phase angle (gibbous phase, solid blue) and another observation is 
           taken half an orbit later (crescent phase, dashed blue).  The gibbous 
           observations occur in the middle of northern summer while the crescent 
           observations occur in the middle of northern winter.  The system is 
           assumed to be viewed edge on (inclination of 90$^{\circ}$) in the 
           top row (where gibbous and crescent phase observations actually refer 
           to full and new phase, respectively), and is viewed at an inclination 
           of 60$^{\circ}$ in the bottom row.  In the ``No Moon" cases, the 
           difference between gibbous and crescent observations (black line) shows 
           only seasonal variability, which is very small in the 4.3~$\mu$m carbon 
           dioxide band and the 6.3~$\mu$m water band.  For the observations in 
           which the Moon is present, these bands are filled in by the lunar flux 
           at gibbous phase, and the difference between the gibbous and crescent 
           observations shows much larger variability within the absorption bands.}
  \label{fig:diffs}
  \label{lastfig}
\end{figure}

In Fig.~\ref{fig:diffs} we demonstrate the differencing approach. 
An exoEarth (left column, ``No Moon") as well as a exoEarth-Moon 
system (right column, ``Moon") are observed at a distance of 10~pc at 
an inclination of 90$^{\circ}$ (edge on) and 60$^{\circ}$.  The 
observations are averaged over 24~hours in either mid-northern 
summer or winter.  One observation is taken at the smallest possible 
phase angle, which is determined by the inclination, and another 
observation is taken half an orbit later when the planet/system 
are at the largest possible phase angle.  The difference between 
these two observations shows only seasonal variability in the 
Earth-alone case, and shows a combination of the variability from 
seasons and the Moon in the Earth-Moon case.  Without the presence of 
the Moon, variability within the 4.3~$\mu$m carbon dioxide band and 
the 6.3~$\mu$m water band is quite small; on its own, Earth's spectrum 
is both dark and stable within these bands.  However, when the phase 
dependent lunar flux is included, variability in these bands is much 
larger, and the difference between the small phase angle observation and 
the large phase angle observation closely resembles the small phase angle 
contribution from the Moon at these wavelengths.  Thus, variability 
within the 4.3~$\mu$m carbon dioxide band and the 6.3~$\mu$m water is an 
indicator of the presence of a moon.


We investigate the differencing approach for a wider range of 
planetary system inclinations and summarize the results 
in Table I.  Except for inclination, the system parameters 
are the same as in the previous paragraph.  The exoEarth-Moon 
system is observed at the smallest possible phase angle in 
northern summer and at the largest possible phase angle in 
northern winter, and the observer is placed over the northern 
hemisphere.  Note that system inclination affects the range of 
possible phase angles that can be observed, and that an 
inclination of 0$^{\circ}$ corresponds to viewing the system 
face on.  Bands spanning 4.2-4.5~$\mu$m and 5.0-7.5~$\mu$m 
(in the 4.3~$\mu$m carbon dioxide band and 6.3~$\mu$m water 
band, respectively) were found to be ideal for detecting the 
lunar signal, where a balance must be achieved between a wide 
enough band for photon collection and a narrow enough band 
to exclude seasonal variability outside the absorption 
feature.  In addition, Table I shows flux ratios for the 
exoEarth-Moon system, which demonstrates the significance of 
the exoMoon's brightness at some wavelengths, as well as inferred 
brightness temperatures for the Earth twin (assuming the 
observer is ignorant of the contamination by, and presence of, 
the companion).

Table I also shows 
an estimate of the minimum required SNR for the gibbous and 
crescent phase observations such that their difference would 
measure the gibbous phase lunar flux at a SNR of 10.  By simple 
error propagation, this SNR is given by 
$SNR^{\prime}\sqrt{F_{G}^{2}+F_{C}^{2}}/(F_{G}-F_{C})$, 
where $SNR^{\prime}$ is the SNR for the measurement of the gibbous 
phase lunar flux (which we take to be 10), and $F_{G}$ and $F_{C}$ 
are the gibbous and crescent phase fluxes, respectively, for the 
planet-moon system through a given bandpass.  Table I shows SNR 
estimates for detections in the 4.3~$\mu$m carbon dioxide band and 
the 6.3~$\mu$m water band for a Moon twin and for a body twice the 
size of the Moon.

In the 4.2-4.5~$\mu$m range, the gibbous phase flux from the 
exoMoon is more than 300\% larger than the exoEarth's  
variability for a wide range of phases.  As a result, the SNR 
required to detect the exoMoon's thermal signal is rather 
small (between 10-20) for all inclinations above 
30$^{\circ}$.  At inclinations below about 30$^{\circ}$, the 
crescent phase lunar flux is a sizeable fraction of the gibbous 
phase flux, causing the crescent phase flux to contaminate the 
measurement of the gibbous phase lunar flux when subtracting 
the observations taken at different phases.  In the 5.0-7.5~$\mu$m 
range, the exoEarth's variability begins to wash out the 
gibbous phase thermal flux from the exoMoon at inclinations 
below about 45$^{\circ}$. At inclinations above this, the required 
SNR for detection is only slightly larger than 20, and is close to 
10 for companions twice the size of the Moon. In general, 
contamination from the exoEarth's seasonal variability and the 
crescent phase lunar signal cause the differencing technique to work 
poorly for inclinations below about 30-45$^{\circ}$. For inclinations 
above this, detecting Moonlike satellites via the differencing 
technique is feasible.

\section{Discussion}

Surface, tropospheric, and stratospheric temperatures on Earth 
are typically within the range of 200-300~K, and drastic 
day/night temperature differences do not occur due to atmospheric 
circulation as well as relatively large surface and atmospheric 
heat capacities.  The Moon, in contrast, has a relatively low 
surface heat capacity and lacks an atmosphere with which to 
redistribute energy from the day side to the night side of the 
world.  As a result, surface temperatures on the Moon are as high 
as 400~K at the sub-solar point, allowing the Moon to contribute a 
significant amount of flux to IR observations of the Earth-Moon 
system (depending on phase).  Furthermore, the large lunar 
day side temperatures cause the peak of the lunar thermal 
spectrum to be located at wavelengths distinct from the peak of 
Earth's thermal spectrum.  Near full phase, the peak of the 
lunar thermal spectrum occurs near the 6.3~$\mu$m water band, 
causing the Moon to outshine Earth both at these wavelengths as 
well as in the 4.3~$\mu$m carbon dioxide band.

When observing an unresolved Earth-Moon system, thermal flux 
from the Moon disproportionately affects regions of Earth's 
spectrum where Earth has strong absorption bands.  As a result, 
characterization of Earth's atmospheric composition and 
temperature from IR observations taken by a distant 
observer could be strongly influenced by the Moon.  For example, 
for full phase observations of the Earth-Moon system, lunar 
thermal radiation consistently comprises about 20\% of the total 
signal, makes up as much as 80\% of the total signal in the 
6.3~$\mu$m water band (creating the appearance of a much drier 
planet), and over 90\% of the signal in the 4.3~$\mu$m carbon 
dioxide band. Current models predict that large impacts like 
the Moon forming impact should be common and that conditions 
present in the debris disk following such an impact cause 
any companions formed from debris material to be depleted in 
volatiles.  Thus, contamination of IR observations of extrasolar 
terrestrial planets due to unresolved, airless companions may be 
a common reality, and the discussion in the previous paragraphs 
will generally apply to thermal observations taken by future exoplanet 
detection and characterization missions.  

It is important to point out that contamination from airless 
companions can be minimized by taking observations at the largest 
feasible phase angles.  Such a configuration maximizes the flux 
from the companion's cold night side, and minimizes flux from the 
warmer day side.  However, depending on the orbital inclination of the 
system, large phase angles may not be accessible.  In this case, the 
contribution from the companion will be nearly constant and relatively 
small, except in some absorption bands (Fig.~\ref{fig:examples}, second 
panel).

The contribution from an airless companion depends strongly on 
phase angle and, thus, can mimic seasonally dependent thermal 
variations from the host planet.  Figure~\ref{fig:diffs} 
demonstrates how drastic this effect can be for Earth and 
the Moon.  An exoEarth-Moon system was observed at 
gibbous phase in the middle of northern summer so that the lunar 
signal adds to the seasonal variability in the exoEarth's  
spectrum, causing the flux variations in the atmospheric window 
region to appear roughly twice as large as the Earth-only case.  If 
the presence of the companion goes undetected, then it appears as 
though the exoEarth has very exaggerated seasons.  If the gibbous 
observation were to occur instead in the middle of northern winter, 
then the lunar contribution would wash out the seasonal variations from 
the exoEarth, and the variability in the atmospheric window region 
would decrease to nearly zero. This would create the false appearance 
of a planet with almost no seasonal climate variability. These findings 
further demonstrate how the presence of an undetected companion can 
interfere with the measurement of the obliquity and thermal properties 
of the host planet, and are in good agreement with the bolometric 
results in \citet{moskovitzetal09}.

The ability of a companion to an extrasolar terrestrial planet 
to outshine its host at some wavelengths proves to be useful  
as the phase dependent variability in the companion's 
brightness can impart a detectable signal in spatially 
unresolved observations of the planet-companion system.  The 
broadband models of \citet{moskovitzetal09} did not capture  
this important behavior, causing them to conclude that only 
large satellites (roughly Mars sized) of Earthlike exoplanets 
could be detected by NASA's Terrestrial Planet Finder (TPF).  As 
shown in Table I, it is actually quite feasible to detect exomoons 
by differencing gibbous phase and crescent phase observations 
of the planet-moon system at wavelengths where the moon 
is bright and the planet's spectrum is relatively dark and stable.  
A band spanning 4.2-4.5~$\mu$m is well suited to detecting exomoons 
as the flux from an exoEarth is quite small and stable in this region, 
and variability from an exoMoon could be detected with observations 
taken at a SNR of about 10. Note that the SNRs presented in Table I 
are close to or within the capabilities of current TPF strategies, but 
that current TPF-Interferometer science requirements use a shortwave 
cutoff at 6.5~$\mu$m, which does not reach the bottom of the 
6.3~$\mu$m water band or the 4.3~$\mu$m carbon dioxide 
band \citep{beichmanetal06}.

The phase differencing technique outlined in this paper should 
function generally for terrestrial exoplanets and their airless 
moons provided that there exists a wavelength range where the moon 
contributes a significant amount of the gibbous phase flux from 
the unresolved system and where the flux from the host planet is 
relatively stable.  The peak in the gibbous phase Moon's spectrum 
is at about 7~$\mu$m, which is near the 4.3~$\mu$m carbon dioxide 
band and the 6.3~$\mu$m water band, where the escaping flux from 
Earth is coming from cold regions of the atmosphere near the upper 
troposphere and lower stratosphere, at which temperature variability 
is quite small as compared to surface temperature variations.  Searching 
for excesses due to an exomoon in the 4.3~$\mu$m carbon dioxide band 
is attractive since CO$_{2}$ is a common well mixed gas in terrestrial 
planetary atmospheres.  Using the 6.3~$\mu$m water band will be useful 
for habitable exoplanets which, almost by definition, will present a 
deep water band and whose moons will be receiving a stellar flux 
similar to what the Moon receives, heating these companions to 
temperatures similar to our Moon.  The 7.7~$\mu$m methane band 
would be highly suitable for detecting thermal excesses from moons 
orbiting planets analogous to the early Earth, which was expected 
to have much higher atmospheric methane concentrations than the 
modern Earth \citep{kastingetal01}.

Interesting investigations for the future include pairing our models 
with reverse/retrieval models for terrestrial exoplanets to further 
explore the extent to which an exomoon could confound spectroscopic 
characterization of an exoEarth.  Our Earth model is dependent on 
input data from Earth observing satellites, so that we cannot apply 
the current model to terrestrial planets with seasonal cycles different 
from those on Earth. However, pairing our spectral model of Earth 
to a three-dimensional general circulation model for Earthlike planets is 
a task that would enable us to model time dependent, high resolution 
spectra of terrestrial planets with distinct climates from Earth.  Such a 
study would allow us to understand whether or not the 6.3$~\mu$m water band 
is stable enough to allow for the detection of exomoons around planets with 
high obliquity angles or planets with eccentric orbits (\emph{i.e.}, 
planets with more extreme seasons than Earth).

\section{Conclusions}

Depending on viewing geometry, the Moon can contribute a significant 
amount of flux to IR observations of a spatially unresolved 
Earth-Moon system, especially at wavelengths where there are strong 
absorption bands in Earth's spectrum.  For an extrasolar Earth-Moon 
system observed at full phase, the Moon consistently comprises about 
20\% of the total signal at most wavelengths, and makes up as much as 
80-90\% of the total signal in the 6.3~$\mu$m water band and the 
4.3~$\mu$m carbon dioxide band.  The added flux in the water band 
creates the appearance of a more desiccated planet, resembling the 
spectrum of Earth with atmospheric water vapor mixing ratios 
artificially lowered to 10\% of their present values.  Furthermore, 
the added lunar flux can increase inferred brightness temperatures for 
Earth by as much as 40~K at some wavelengths.  Thermal flux from an 
airless exomoon depends strongly on phase angle, so that differencing 
observations taken at small phase angles from those taken at large phase 
angles, at wavelengths where the host planet's spectrum is relatively 
stable over seasonal timescales, it may be possible to detect the excess 
thermal radiation coming from a Moon-sized companion in the gibbous phase 
observations using a TPF-like telescope.

\acknowledgements

This work was performed as part of the NASA Astrobiology Institute's 
Virtual Planetary Laboratory, supported by the National Aeronautics 
and Space Administration through the NASA Astrobiology Institute under 
solicitation No.~NNH05ZDA001C.  I would like to thank Vikki Meadows, 
Eric Agol, David Crisp, and Josh Bandfield for discussions and insights 
provided throughout the course of this project. Some of the results in 
this paper have been derived using the HEALPix \citep{gorskietal05} package.

\clearpage

\section{Tables and Figures}

\begin{landscape}
\begin{table}[htbp]
 \caption{ {\scriptsize Phase Differencing Technique for Detecting Exomoons: Thermal Fluxes, Flux Ratios, Brightness Temperatures, and Estimated SNR Requirements} }
 \begin{center}
 {\tiny
  \begin{tabular}{ c || c | c | c | c | c | c | c | c | c | c | c | c || c | c }
     \hline\hline
                      & \multicolumn{12}{c||}{ Flux at 10 pc, exoMoon/exoEarth Flux Ratio ($F_{M}/F_{E}$), and Brightness Temperature\footnotemark[1] ($T_{b}$)}                                                                                                  & \multicolumn{2}{c}{ SNR for Detection\footnotemark[2] }            \\
                      & \multicolumn{12}{c||}{ Gibbous (Crescent) }                                                                                                                                                                                               & \multicolumn{2}{c}{                                   }            \\
                      & \multicolumn{4}{c|}{ Bolometric }                                               & \multicolumn{4}{c|}{ 4.2 - 4.5~$\mu$m }                                          & \multicolumn{4}{c||}{ 5.0 - 7.5~$\mu$m }                                         & \multicolumn{2}{c}{  4.2-4.5~$\mu$m (5.0-7.5~$\mu$m)  }\\
                      & \multicolumn{2}{c|}{ Flux / $10^{-20}$ [W/m$^{2}$]} & $F_{M}/F_{E}$ & $T_{b}$   & \multicolumn{2}{c|}{ Flux / $10^{-23}$ [W/m$^{2}$] } & $F_{M}/F_{E}$ & $T_{b}$   & \multicolumn{2}{c|}{ Flux / $10^{-21}$ [W/m$^{2}$] } & $F_{M}/F_{E}$ & $T_{b}$   & \multicolumn{2}{c}{                                   }\\ 
     Inc [$^{\circ}$] & Earth              & Moon                           & [\%]          & [K]       & Earth              & Moon                            & [\%]          & [K]       & Earth              & Moon                            & [\%]          & [K]       & $R_{M}$ & 2$R_{M}$                                     \\ \hline\hline
     90               & 10.6 (9.6)         & 2.6 ($<$0.1)                   & 24 ($<$1)     & 272 (251) & 3.0 (2.4)          & 14.1 ($<$0.1)                   & 480 ($<$1)    & 272 (234) & 3.6 (3.2)          & 3.8 ($<$0.1)                    & 110 ($<$1)    & 266 (241) & 12 (22) & 11 (13)                                      \\
     75               & 10.6 (9.3)         & 2.5 ($<$0.1)                   & 23 ($<$1)     & 271 (249) & 3.0 (2.2)          & 13.7 ($<$0.1)                   & 450 ($<$1)    & 271 (232) & 3.6 (3.0)          & 3.7 ($<$0.1)                    & 100 ($<$1)    & 266 (240) & 12 (21) & 11 (13)                                      \\
     60               & 10.5 (8.9)         & 2.3 (0.1)                      & 21 (1)        & 270 (247) & 3.2 (2.0)          & 12.3 ($<$0.1)                   & 390 (2)       & 270 (231) & 3.6 (2.9)          & 3.3 ($<$0.1)                    & 90 (1)        & 264 (239) & 13 (23) & 11 (13)                                      \\
     45               & 10.5 (8.6)         & 1.9 (0.1)                      & 18 (2)        & 267 (245) & 3.2 (1.8)          & 10.2 (0.2)                      & 320 (10)      & 266 (231) & 3.5 (2.7)          & 2.8 (0.1)                       & 80 (4)        & 261 (238) & 14 (25) & 11 (14)                                      \\
     30               & 10.5 (8.4)         & 1.6 (0.3)                      & 15 (4)        & 265 (245) & 3.3 (1.7)          &  7.8 (0.7)                      & 240 (40)      & 262 (234) & 3.5 (2.6)          & 2.2 (0.3)                       & 60 (1)        & 258 (239) & 16 (33) & 12 (17)                                      \\
     15               & 10.5 (8.4)         & 1.2 (0.5)                      & 11 (6)        & 263 (246) & 3.3 (1.7)          &  5.4 (1.7)                      & 170 (100)     & 257 (240) & 3.5 (2.6)          & 1.6 (0.6)                       & 50 (2)        & 255 (241) & 25 (60) & 18 (28)                                      \\
      0               & 10.5 (8.5)         & 0.8 (0.8)                      & 8 (8)         & 261 (249) & 3.2 (1.8)          &  3.3 (3.3)                      & 100 (180)     & 252 (247) & 3.5 (2.7)          & 1.0 (1.0)                       & 30 (4)        & 251 (245) &         &                                              \\ \hline\hline
  \end{tabular}}
 \end{center}
 \label{tab:diffmeth}
 \label{lasttable}
\end{table}
\footnotetext[1]{Brightness temperatures are computed using the net flux from 
the system assuming a size of one Earth radius in the conversion from flux to 
intensity.}
\footnotetext[2]{A ``detection" constitutes measuring the excess gibbous phase 
lunar flux at a SNR of 10, which is accomplished by differencing the gibbous and 
crescent phase observations of the system.  Estimates of the required SNR are shown 
for a body with a radius equal to the Moon's radius ($R_{M}$), and for a body twice 
as large as the Moon. SNR calculations are further described in the text.}
\end{landscape}

\clearpage


\end{document}